\documentclass[a4,11pt]{article}
\usepackage{epsfig}

\usepackage{amssymb,epsfig,graphicx,epsf}

\usepackage{amsfonts}
\usepackage{amsmath}
\usepackage{amssymb}
\usepackage{feynmf}
\newcommand{\bee}{\begin{equation}}
\newcommand{\ee}{\end{equation}}

\newcommand{\ba}{\begin{eqnarray}}
\newcommand{\ea}{\end{eqnarray}}
\newcommand{\bal}{\begin{aligned}}
\newcommand{\eal}{\end{aligned}}

\begin{document}

\title{Gravitational waves in the presence of a cosmological constant}

\author{Jos\'e Bernabeu\\
Departament de F\'\i sica Te\`orica and IFIC, Universitat de Val\`encia-CSIC,\\ 
Dr. Moliner 50,  46100 Burjassot, Val\`encia, Spain.\\
\\
Dom\`enec Espriu and Daniel Puigdom\`enech\\
Departament d'Estructura i Constituents de la Mat\`eria,\\ 
Institut de Ci\`encies del Cosmos (ICCUB)\\ Universitat de Barcelona\\
Mart\'\i ~i Franqu\`es, 1, 08028 Barcelona, Spain.}

\date{}

\maketitle

\begin{abstract}
We derive the effects of a non-zero cosmological constant $\Lambda$ on gravitational wave propagation
in the linearized approximation of general relativity. In this approximation we consider the situation where the metric 
can be written as $g_{\mu\nu}= \eta_{\mu\nu}+ h_{\mu\nu}^\Lambda + h_{\mu\nu}^W$, $h_{\mu\nu}^{\Lambda,W}<< 1$, where  $h_{\mu\nu}^{\Lambda}$ is the background perturbation and  $h_{\mu\nu}^{W}$ is a modification interpretable as a gravitational wave.
For $\Lambda \neq 0$ this linearization of Einstein equations is self-consistent only in
certain coordinate systems. The cosmological Friedmann-Robertson-Walker 
coordinates do not belong to this class and the derived linearized solutions have to be 
reinterpreted in a coordinate system that is homogeneous and isotropic to make contact with observations. 
Plane waves in the linear theory acquire modifications of order $\sqrt{\Lambda}$, both in the amplitude and 
the phase, when considered in FRW coordinates. In the linearization process for $h_{\mu\nu}$, we have also included terms of order $\mathcal{O}(\Lambda h_{\mu\nu})$. For the background perturbation  $h_{\mu\nu}^\Lambda$ the difference is very small but when the term $h_{\mu\nu}^{W}\Lambda$ is retained the equations of motion can 
be interpreted as describing massive spin-2 particles. However, the extra degrees of freedom can be approximately 
gauged away, coupling to matter sources with a strength proportional to the cosmological constant 
itself. Finally we discuss the viability of detecting the modifications caused by the cosmological 
constant on the amplitude and phase of gravitational waves. In some cases the distortion with
respect to gravitational waves propagating in Minkowski space-time is considerable. The effect of $\Lambda$ could have a detectable impact on pulsar timing arrays.
\end{abstract}

\vfill
\noindent
June 2011

\noindent
FTUV-11-2106

\noindent
UB-ECM-FP-55/11

\noindent
ICCUB-11-151

\section{Introduction}
The smallness of the cosmological constant obtained from fits to the current $\Lambda$CDM 
cosmological models\cite{concordance} ($\Lambda\simeq 10^{-52}$ m$^{-2}$) may lead us to 
believe that it is totally unobservable except at the largest distances. However, the issue 
of the relevance of the cosmological constant 
in local measurements (meaning measurements that involve
sub-cosmological scales, such as for instance galaxy clusters) has received growing attention\cite{previous,ber}. 
One interesting possibility is assessing the influence of $\Lambda$ on the bending 
of light from distant objects. At present there are rather diverging results on the 
subject giving rather different results concerning the relevance of $\Lambda$ ranging from
zero\cite{khriplovich} or very small\cite{sereno} to appreciable ones\cite{rindler1}. 
The effect of $\Lambda$ on the photon propagation, including frequency shift, Shapiro time delay and deflection of light, is 
currently under consideration\cite{bmv}. 

The importance of these studies cannot be overemphasized. The presence of a non-zero cosmological 
constant contributing around 70\% to the energy and matter budget of the universe, 
seemingly making the Universe globally a de Sitter space-time, is one of 
the intriguing puzzles of Physics in our time. Observations capable 
of confirming or refuting the relevance of $\Lambda$ at redshift $z < 1$ 
are clearly of utmost importance. 

The studies of what has been termed `local gravity with a cosmological constant' rely on an approximate solution, 
valid at first order in $\Lambda$, obtained after linearizing Einstein equations. These solutions have recently
been studied in detail by one of the authors\cite{ber} using different gauge choices. 
It has been found that in the Lorenz gauge one can in addition require time independence of the metric solutions. After an additional 
coordinate transformation these solutions correspond to the linearized version of the Schwarzschild-de Sitter
exact solution of Einstein equations. The modification to the Newtonian limit in such coordinates was
also discussed in detail\cite{ber}. There are some subtleties related to
the physical interpretation of the different coordinate systems that we shall review below.   

Here we propose to study a different problem. Namely, how $\Lambda$ influences the properties
of gravitational waves (GW). As of today, gravitational waves are an unambiguous prediction of General Relativity
that has not been tested directly. They are `observed' indirectly as they are the missing ingredient   
needed to restore the energy balance of some astrophysical binary systems\cite{binary}. 
There are three types of experiments
potentially capable of yielding a non-zero signal in the coming years. Let us summarize their physical
and astrophysical reach here:

Ground based GW detectors such as LIGO \cite{ligo} can reach sensitivities down to 
$\sim 10^{-23}$  with optimal sensitivity in the region between 10 Hz and $10^3$ Hz.
The space mission LISA\cite{lisa} will reach a similar sensitivity in the range
$10^{-2} $ Hz to $10^{-3}$ Hz but will actually be able to set relevant bounds on a more extended range
of frequencies. Finally the International Pulsar Timing Array project\cite{IPTA}
or the Square Kilometer Array project\cite{SKA} are sensitive to lower frequencies 
$\nu < 10^{-4}$ Hz but reach only a sensitivity of $\sim 
10^{-10}$ going up to  $\sim 
10^{-15} $ for $\nu \sim 10^{-10}$ Hz. These sensitivity ranges are targeted to
specific astrophysical phenomena and are expected to provide detectable signals and confirm
the existence of GW in the coming decades. 

Given the present difficulties in asserting the very existence of GW it may seem academic to try to find 
modifications due to the presence of a cosmological constant that is small. However, it should 
be borne in mind that in the inflationary epoch the value
of $\Lambda$ was much larger than at present so these effects might be of relevance for
primordial GW. As we will discuss in this work the effect of $\Lambda$ could be
of some relevance for GW travelling very long distances and for pulsar timing array projects. On the other hand, some of 
the results presented here we believe are of interest to
understand the issue of the gauge choice in the presence of $\Lambda$ for the linear theory. Finally, it seems interesting
in its own right to attempt to understand wave propagation in de Sitter space-time if 
$\Lambda$ is indeed a fundamental parameter of nature.          

This paper is organized as follows. In section 2 we discuss the linearization of 
Einstein equations, including a discussion on different gauges and how they affect the
wave equation for the gravitational field $h_{\mu\nu}$. In section 3 we discuss different coordinate realizations of de Sitter space-time and their relation. 
In section 4 we construct background solutions retaining terms of order 
$\Lambda h_{\mu\nu}$. This discussion is extended in section 5 to include GW solutions
that `feel' the presence of $\Lambda$. In section 6 we analyze the detectability 
of the effects previously calculated. In section 7 we summarize the conclusions of this study.

Some of the subjects discussed here appear to have received little attention in the past although 
there is an extensive literature on gravitational waves \cite{gwsummary}. The effect of $\Lambda$
on GW has been considered in \cite{bic,expansion}. Physical consequences appear
to have been extracted in the context of primordial gravitational waves\cite{bsti} and only indirectly
in what concerns the evolution of the modes and the power spectrum.

\section{Linearization in the presence of $\Lambda$}
Einstein equations, derived from the Einstein-Hilbert action, read
\bee \label{1}
R_{\mu\nu}-\frac{1}{2} g_{\mu\nu}R+\Lambda g_{\mu\nu}=-\kappa T_{\mu\nu}
\ee
where $R_{\mu\nu}$ is the Ricci tensor for $g_{\mu\nu}$, $\Lambda>0$ is the cosmological constant 
and $\kappa T_{\mu\nu}$ is the source term. $T_{\mu\nu}$ is the usual stress-energy tensor of matter 
in the gravitational field generated by $g_{\mu\nu}$ and $\kappa$ is the dimensionful constant 
coupling matter and gravity. However, throughout this work we will consider $T_{\mu\nu}=0$ unless 
otherwise specified. The inclusion of the cosmological constant term leads to curvature even 
in the absence of any source
\bee \label{2}
R=4\Lambda .
\ee
We consider the linearized theory where the metric is written as
\bee \label{3}
g_{\mu\nu}=\eta_{\mu\nu}+h_{\mu\nu},
\ee
$\eta_{\mu\nu}$ being the Minkowski metric and $h_{\mu\nu}<< 1$. The Ricci tensor to first order in the small perturbation $h_{\mu\nu}$ reads
\bee \label{4}
R_{\mu\nu}=\frac{1}{2}\left(\Box h_{\mu\nu}+h_{,\mu\nu}-h_{\mu ,\nu\lambda}^{\lambda}-
h_{\nu ,\mu\lambda}^{\lambda}\right),
\ee
indices being lowered and raised with $\eta_{\mu\nu}$ and $h=\eta^{\mu\nu}h_{\mu\nu}$. 
The theory is invariant under coordinate transformations $x^\mu \to x^\mu + \xi^\mu(x)$. 
For infinitesimal transformations the perturbation metric
$h_{\mu\nu}$ transforms as $h_{\mu\nu}\to h^\prime_{\mu\nu}= 
h_{\mu\nu}+ \partial_\mu\xi_\nu+\partial_\nu\xi_\mu$. 
A gauge choice is possible, amounting to selecting a particular class of coordinates, and in fact 
such a choice is necessary if the perturbation $h_{\mu\nu}$ is to be quantized. 
In order to discuss GW two different gauge choices are particularly appropriate.

\subsection{Lorenz gauge}

In order to describe perturbations around flat space-time it is customary to employ the Lorenz gauge.
\bee
\partial_{\mu}h_{\,\nu}^{\mu}=\frac{1}{2}\partial_{\nu}h,\label{lg}
\ee 
or
\bee
\partial_{\mu}\tilde h_{\,\nu}^{\mu}=0,
\ee
where
\bee
\tilde h _{\mu\nu} = h_{\mu\nu}-\frac12 \eta_{\mu\nu} h
\ee
is the trace reversed version of $h_{\mu\nu}$.

In this gauge, expression (\ref{4}) is simplified
\bee \label{5}
R_{\mu\nu}=\frac{1}{2}\Box h_{\mu\nu},
\ee
and we obtain the equation of motion
\bee\label{6}
\Box \left(h_{\mu\nu}-\frac{1}{2}\eta_{\mu\nu}h\right) + 2\Lambda h_{\mu\nu}= -2\Lambda\eta_{\mu\nu}
\ee
which has always to be considered together with the Lorenz gauge condition (\ref{lg}). 

Whether the term of order $\mathcal{O}(h\Lambda)$ has to be considered or not depends on the relative magnitude 
of $h$ and $\Lambda$. There will be situations where the inclusion of this term is justified and may lead to 
observable consequences. We shall postpone the rest of the discussion on this issue to sections 4 and 5. 
Note, nonetheless, that if the $\Lambda h_{\mu\nu}$ term on the l.h.s. is omitted (and only in this case)
there is a residual gauge freedom within the Lorenz gauge. 
If we perform a linear coordinate transformation
\bee \label{5b}
x^{\mu}\rightarrow x'^{\mu}=x^{\mu}+\xi^{\mu},
\ee
equation (\ref{lg}) is fulfilled as long as $\xi^{\mu}$ is an harmonic function, i.e. $\Box \xi^{\mu}=0$. 
These residual coordinate transformations are sometimes termed 
`coordinate waves' for rather obvious reasons. Note also that whether this 
is a symmetry of the equations of motion or not, depends on the terms retained in the linearization; 
the term $\Lambda h_{\mu\nu}$ breaks this residual coordinate invariance.

\subsection{$\Lambda$ gauge}

It will be useful to consider an alternative gauge choice\cite{lu}, which we 
will term $\Lambda$-gauge. This is given by the gauge condition
\bee
\partial_{\mu}\tilde{h}_{\nu}^{\mu}=-\Lambda \eta_{\nu\mu}x^{\mu}.\label{lamg}
\ee
In this gauge the linearized equations of motion look slightly different
\bee\label{8}
\Box \left(h_{\mu\nu}-\frac{1}{2}\eta_{\mu\nu}h\right) - 2\Lambda h_{\mu\nu}= 0.
\ee
In particular we note that the term independent of $h_{\mu\nu}$ on the r.h.s.
of (\ref{6}) is absent. There is a set of coordinate transformations 
that can be performed without leaving the gauge orbit (\ref{lamg}); these are 
transformations $ x'^{\mu}=x^{\mu}+\xi^{\mu}$ with
\bee \label{9.5}
\Box \xi^{\mu}= -\Lambda \xi^\mu. 
\ee
However, in the $\Lambda$-gauge this residual coordinate transformations are not a 
symmetry of the equations of motion regardless of the terms retained in the linearization 
and therefore cannot be used to remove degrees of freedom. Generally speaking, linearization leaves global 
Lorentz transformations as the only symmetry of the equations of motion. The Lorenz gauge
is in a way special as some additional freedom to perform local coordinate
transformations remains if the term $\Lambda h_{\mu\nu}$ is neglected. The situation in the 
$\Lambda$-gauge, where there is no residual symmetry, is, on the contrary, the generic one.

The connection between the two gauge choices in the linear theory is 
easily made when the terms $\Lambda h_{\mu\nu}$ are omitted. 
It is implemented via the following change of coordinates
\bee \label{9}
x^{\mu} \rightarrow x'^{\mu}=x^{\mu}+\xi^{\mu}=\left(1-\frac{\Lambda}{12}x^2\right)x^{\mu}.
\ee
This change of coordinates transforms a solution of $\Box \tilde h_{\mu\nu}=0$ in the $\Lambda$-gauge 
(coordinates $x$) to a solution of $\Box \tilde h_{\mu\nu}= -2\Lambda \eta_{\mu\nu}$ in Lorenz gauge
(coordinates $x^\prime$). Note the simplicity of the equation for linear perturbations in
the $\Lambda$-gauge if the term of order $\Lambda h_{\mu\nu}$ is omitted. All reference to
the cosmological constant is eliminated.

The previous discussion in the $\Lambda$-gauge reminds us that in general, in the linearized 
approximation, the perturbation metric $h_{\mu\nu}$ is expected to have up to six full degrees of 
freedom. Only in certain cases a residual gauge freedom 
can be used to further reduce the number of degrees of freedom.

Let us elaborate a bit more on this issue as it is conceptually important. In the Lorenz gauge, 
with the term $\Lambda h_{\mu\nu}$ omitted, the residual symmetry (\ref{5b}) 
allows us to move freely between different coordinate systems, say $x^\prime$ and $y^\prime$, which are not
trivially related by Lorentz transformations and yet preserve
the form of the equations of motion. On the contrary, if we undo transformation (\ref{9}) we get two coordinate 
systems $x$ and $y$ in which the $\Lambda$-gauge condition is fulfilled but at best only one of 
these gauge transformed coordinate systems obeys the linearized equations of motion in 
the $\Lambda$-gauge; the other one is off-shell. That is to say, the number of 
independent degrees of freedom seems to be larger in the $\Lambda$-gauge. However, since this is purely due to
a gauge choice, the additional apparent degrees of freedom cannot correspond to physical ones. 

If the term of order $\Lambda h_{\mu\nu}$ is retained, i.e. in the Lorenz gauge the term $2 \Lambda h_{\mu\nu}$  
on the l.h.s. of (\ref{6}) or the analogous $-2 \Lambda h_{\mu\nu}$ in the $\Lambda$-gauge are kept,
there is no residual symmetry whatsoever. Let us take for example (\ref{6}) in 
the Lorenz gauge; as we will see in detail in section 5 this generates
a genuine mass term and therefore more physical degrees of freedom
appear associated to $h_{\mu\nu}$. This  is not a gauge artifact.

\section{De Sitter space-time}

De Sitter space-time can be described by many coordinate systems. A convenient choice of coordinates is 
Schwarzschild-de Sitter (SdS). These provide a time-independent metric in a gauge that is none 
of the two previously discussed
\bee \label{9.1}
\bal
d{ s}^2=&\left[1-\frac{\Lambda}{3}{\hat r}^2\right]d{\hat t}^2-
\left[1-\frac{\Lambda}{3}{\hat r}^2\right]^{-1}{\hat r}^2+{\hat r}^2d\Omega^2.\\
\eal
\ee
and clearly shows the presence of the de Sitter horizon. We note that this metric admits an expansion
in integer powers of $\Lambda$. Note also that in this metric the spatial part does not quite correspond to
spherical coordinates. 

At the opposite extreme, one can select a metric that depends only on time and is position independent.
It is the Friedmann-Robertson-Walker (FRW) metric
\bee \label{9.2}
ds^2= dT^2 -\exp(2\sqrt{\frac{\Lambda}{3}}T) d\vec X^2.
\ee
This metric incorporates the physical principles of cosmological homogeneity and isotropy as it does 
not depend on the position. The coordinates $X^i$ have a clear physical meaning, they are comoving coordinates
anchored in space that expand with the universe. These are the natural coordinates where our 
world appears homogeneous and isotropic. 
It is easy to see that the FRW metric does not fulfill any linearized Einstein equation, even
for very early times  $t << 1/\sqrt{\Lambda}$ when is very close to the  
Minkowski metric. In fact, no metric that depends only on time can be
a solution of the linearized Einstein equations; incompatibilities appear immediately for any gauge
choice.

One should therefore accept that the linearized Einstein equations in the presence of $\Lambda$ cannot be imposed
in the physically relevant comoving coordinate system. This of course has implications on GW as 
the very concept of `wave' does require a wave equation, which is just impossible in FRW coordinates.
On the other hand, the wave equation $\Box \tilde h_{\mu\nu}=0$ found in the $\Lambda$-gauge
is expressed in a set of coordinates whose meaning is yet to be interpreted. 
Therefore the simplicity of this equation is deceiving.
 
We will argue in the next section that the coordinates implied by the choice of the $\Lambda$-gauge
or of Lorenz gauge are closely related to SdS coordinates. Then the way to proceed is to find 
a solution for GW in the Lorenz gauge, a coordinate system where linearization of the Einstein equations is consistent, and then transform the solution to FRW coordinates in order to extract observable consequences. 

Both the SdS metric and the FRW metric are valid (but rather different) descriptions of de Sitter geometry. 
One can work out the exact transformation between the two coordinate systems
\bee \label{9.3}
\bal
\hat{r}=&e^{T\sqrt{\Lambda /3}}R\\
\hat{t}=&\sqrt{\frac{3}{\Lambda}}\log\left(\frac{\sqrt{3}}{\sqrt{3-\Lambda e^{2T\sqrt{\Lambda /3}}R^2}}\right)+T
\eal
\ee
where $T$ and $R$ are respectively the cosmological time and comoving coordinates whose physical realization 
is clear. This transformation is valid inside the cosmological horizon, i.e. $R<\frac{1}{\sqrt{\Lambda}}$. 
Applying (\ref{9.3}) to (\ref{9.1}) we obtain
\bee \label{9.4}
ds^2= dT^2 -\exp(2\sqrt{\frac{\Lambda}{3}}T) d\vec X^2.
\ee
Now it is immediate to see that the FRW metric does not fulfill any linearized Einstein equation, even if $t<<1/\sqrt{\Lambda}$ as it is not expandable in integer powers of $\Lambda$. The same transformations for the linearized version of the metrics gives
\bee \label{9.5}
\bal
d{s}^2=\left[1-\frac{\Lambda}{3}{\hat r}^2\right]d{\hat t}^2&-
\left[1+\frac{\Lambda}{3}{\hat r}^2\right]{\hat r}^2+{\hat r}^2d\Omega^2.\\
&\downarrow\\
ds^2=dT^2-\left[1+2\sqrt{\frac{\Lambda}{3}}\right. & \left.T+ 2\frac{\Lambda}{3}T^2\right](dR^2+R^2d\Omega^2).\\
\eal
\ee
which will only reasonably approximate the expansion of FRW for values of $R\sim T<<\frac{1}{\sqrt{\Lambda}}$. Note that, although the last metric in (\ref{9.5}) is linearized, it does not fulfill any linearized Einstein equations.

The previous transformation provides the relationship between a framework where the Einstein equations can 
be consistently linearized and the actual coordinate system in which we observe. The solutions easily 
found in the linearized theory have to be transformed to the physically meaningful coordinate system in order to 
make predictions. It is at this point that non-trivial effects related to $\Lambda$ will appear. They
are discussed in section 5. Of course, given the current value of $\Lambda$, these effects will be small. 
We believe nonetheless, that these corrections are conceptually important. Note also that (\ref{9.3}) involves $\sqrt{\Lambda}$ 
and not $\Lambda$, yielding corrections that are potentially much more relevant for 
observation than those of order $\mathcal{O}(\Lambda)$.

Equation (\ref{9.2}) is just one of the many possible cosmological FRW metrics. Other possibilities such as a 
power law cosmological scale factor do not correspond to a de Sitter space-time and therefore there is no 
obvious change of coordinates that allows to reexpress a GW, i.e. a solution to a wave equation, in that 
physically meaningful coordinate system.

\section{Background solutions}
We shall work consistently in 
the linearized approximation both for the background modification
$h_{\mu\nu}^\Lambda$ and for gravitational wave 
perturbations $h_{\mu\nu}^W$. Namely, the metric 
can be written as $g_{\mu\nu}= \eta_{\mu\nu}+ h_{\mu\nu}^\Lambda + h_{\mu\nu}^W$, 
where $h_{\mu\nu}^{\Lambda,W}\ll 1$.
To keep the notation simple we shall only use the superscript $\Lambda$ when confusion 
with wave perturbations $h_{\mu\nu}^W$ is possible. In this section we will be concerned 
with background linearized solutions when the cosmological constant $\Lambda$ is present.

The value of the cosmological constant has presumably not been the same throughout 
the history of the universe. In early epochs, perhaps following an inflationary period, 
its value is believed to have been much larger\cite{inflation}. This fact suggests that it may be
necessary in some circumstances to retain the term $\Lambda h_{\mu\nu}^\Lambda$. 
Likewise it will be necessary for consistency to keep terms of order
$\Lambda h_{\mu\nu}^W$ as the magnitudes of $h_{\mu\nu}^W$ and
$\Lambda$ are unrelated. 

In what follows we proceed without making any assumptions 
on the value of $\Lambda$; we will just assume that the perturbation that induces
on the background metric $h_{\mu\nu}$ is small enough for the linearized approximation to be meaningful.

\subsection{Lowest order solutions}
First we turn to the lowest order solutions already discussed in \cite{ber}, which
correspond to neglecting terms of ${\cal O}(\Lambda h_{\mu\nu})$. In the Lorenz gauge 
this amounts to solving the following equation
\bee \label{10}
\bal
\Box \tilde{h}_{\mu\nu}=&-2\Lambda\eta_{\mu\nu}\\
\partial_{\mu}\tilde{h}_{\nu}^{\mu}=&0.
\eal
\ee
Linearization limits the validity of the solution to values of the coordinates such that $x^2 << 1/ \Lambda$. 

Before discussing the solutions to (\ref{10}) we take a look at the equations in the $\Lambda$-gauge
\bee \label{11}
\bal
\Box \tilde{h}_{\mu\nu}=&0\\
\partial_{\mu}\tilde{h}_{\nu}^{\mu}=&-\Lambda \eta_{\nu\mu}x^{\mu}.
\eal
\ee
Note once more that the linearized equations are not invariant under gauge transformations.
In the Lorenz gauge the cosmological constant is regarded as a gravitational source, it appears 
in the equations of motion, whereas in the $\Lambda$-gauge all dependency in the cosmological constant 
at this order appears through the gauge condition only and in a way it can be interpreted 
as a consequence of the coordinate choice\footnote{
This of course does not mean that the consequences of $\Lambda$ can be removed
by a wise coordinate transformation but it does mean that it disappears from
the equations of motion themselves.}. The connection between the two gauge choices in the linear 
theory has already been discussed.

We can easily solve equations (\ref{11}) to find the traceless solution
\bee \label{12}
\tilde h_{\mu\nu}=-\frac{\Lambda}{18}\left(4x_{\mu}x_{\nu}-\eta_{\mu\nu}x^2\right).
\ee
If we require that the solution is proportional to $\Lambda$ and involves only the
coordinates $x^\mu$ this is the unique solution. In addition, (\ref{12}) is the only one that 
is Lorentz-covariant (note that $\eta_{\mu\nu}$ is the underlying metric and there is
no other four-vector at our disposal).

It is worth noticing that since there is no residual freedom in this gauge, no
transformation can turn this solution into a static metric: \textsl{The $\Lambda$-gauge 
is explicitly incompatible with the solutions being static}. 

We now transform the solution back to the Lorenz gauge using (\ref{9}). We find
\bee \label{13}
h_{\mu\nu}=\frac{\Lambda}{9}\left(x_{\mu}x_{\nu}+2\eta_{\mu\nu}x^2\right).
\ee
Without the $\Lambda h_{\mu\nu}$ term the equation of motion is 
actually invariant under residual transformations. The number of physical
degrees of freedom therefore is reduced to two. This is the only covariant-looking solution in the 
Lorenz gauge but only one of the infinite number of solutions reachable by non-covariant 
residual transformations. The most general form of such transformations is
\bee \label{14}
\xi'^{\mu}=\left(\begin{array}{c}
A(t^2+r^2)t\\
\left(B_{1}t^2+B_{2}x^2+B_{3}(y^2+z^2)\right)x\\
\left(B_{1}t^2+B_{2}y^2+B_{3}(x^2+z^2)\right)y\\
\left(B_{1}t^2+B_{2}z^2+B_{3}(x^2+y^2)\right)z\\
\end{array}\right),
\ee
where $2B_{1}-6B_{2}-4B_{3}=0$. In particular we find the values of these constants that 
allow us to reproduce the static solution of \cite{ber}.
\bee \label{14.1}
A=-\frac{\Lambda}{18},\quad B_{1}=-\frac{\Lambda}{9},\quad B_{2}=-\frac{\Lambda}{18},
\quad B_{3}=\frac{\Lambda}{36};
\ee
One should ask at this point what are these coordinates. We already know that they cannot
correspond to cosmological coordinates. In fact the resulting metric is neither homogeneous nor
isotropic although it preserves the symmetry among the three axes. 
The answer becomes obvious once one discovers that one of the possible residual gauge transformations 
eliminates the time dependence of the metric. A generalization of Birkhoff's 
theorem \cite{theorem} states that there is a unique static solution with spherical symmetry
which is the Schwarzschild-de Sitter metric previously discussed, or more precisely the 
first order of it in the $\Lambda$ expansion. Since Schwarzschild-de Sitter does not fulfill the
Lorenz gauge condition, a time-independent coordinate transformation must also be involved. 
Let us explicitly show this point using a succession of coordinate transformations linear in $\Lambda$.

The first step is to transform (\ref{13}) to a static solution. We start from
\bee \label{41}
\bal
ds^2=&\left[1+\frac{\Lambda}{9}(3t^2-2r^2)\right]dt^2-\left[1-\frac{\Lambda}{9}(-2t^2+2r^2+{x^{i}}^2)\right]{dx^{i}}^2\\
&-\frac{2\Lambda}{9}t~ x^{i}~dt~dx^{i}+\frac{2\Lambda}{9}x^{i}~x^{j}~dx^{i}~dx^{j}
\eal
\ee
where $i=1,2,3$ and $i\neq j$. After the following change of coordinates
\bee \label{42}
\bal
x=&x'+\frac{\Lambda}{9}\left(-t'^2-\frac{x'^2}{2}+\frac{(y'^2+z'^2)}{4}\right)x'\\
y=&y'+\frac{\Lambda}{9}\left(-t'^2-\frac{y'^2}{2}+\frac{(x'^2+z'^2)}{4}\right)y'\\
z=&z'+\frac{\Lambda}{9}\left(-t'^2-\frac{z'^2}{2}+\frac{(x'^2+y'^2)}{4}\right)z'\\
t=&t'-\frac{\Lambda}{18}(t'^2+r'^2)t'
\eal
\ee
the metric transforms into the static solution to order $\Lambda$ found in \cite{ber},
\bee \label{43}
\bal
ds^2=&\left[1-\frac{\Lambda}{3}r'^2\right]dt'^2-\left[1-\frac{\Lambda}{6}(r'^2+3x_{i}'^2)\right]dx_{i}'^2.\\
\eal
\ee
Note that this solution is still in the Lorenz gauge; we only performed a residual gauge transformation
 that is allowed in this gauge. Since our starting solution is only valid to order $\Lambda$, in 
any change of coordinates, either exact or linear, we only keep terms linear in the cosmological constant. 
We can further transform (\ref{43}) to obtain a fully spherically symmetric solution. Under the following change
\bee \label{44}
\bal
x'=&x''+\frac{\Lambda}{12}x''^3\\
y'=&y''+\frac{\Lambda}{12}y''^3\\
z'=&z''+\frac{\Lambda}{12}z''^3\\
t'=&t'',
\eal
\ee
we obtain
\bee \label{45}
\bal
ds^2=&\left[1-\frac{\Lambda}{3}r''^2\right]dt''^2-\left[1-\frac{\Lambda}{6}r''^2\right](dr''^2+r''^2d\Omega^2),\\
\eal
\ee
which does not obey (\ref{10}) anymore. We can now perform another coordinate transformation 
to obtain the SdS metric to order $\Lambda$
\bee \label{46}
\bal
r''=&\hat{r}+\frac{\Lambda}{12}\hat{r}^3\\
t''=&\hat{t}
\eal
\ee
\bee \label{47}
\bal
d{s}^2=&\left[1-\frac{\Lambda}{3}\hat{r}^2\right]d\hat{t}^2-\left[1+\frac{\Lambda}{3}\hat{r}^2\right]d\hat{r}^2+\hat{r}^2d\Omega^2.\\
\eal
\ee
This is the linearized Schwarzschild-deSitter metric. Essentially the background solution  
(\ref{13}) is the SdS metric in a set of coordinates related to SdS by time independent transformations.

\subsection{Next-order solutions} 
Let us now relax the approximation of the previous section and retain terms proportional to $\Lambda h_{\mu\nu}$. 
In particular we will be interested later in terms of order $\Lambda h_{\mu\nu}^W$ that will 
influence the propagation of gravitational waves.

In the Lorenz gauge this requires the simultaneous fulfillment of the two sets of equations (\ref{lg}) and (\ref{6}).
We note that because of the dimensionality of $\Lambda$ any solution of the previous equations 
containing $\Lambda$ and constructed with the only available (Lorentz-)covariant vector $x^\mu$ must
necessarily be even under a change of sign of all coordinates $x^\mu \to -x^\mu$. Solutions 
odd in $x^\mu$ exist but they require the involvement of parameters other than the coordinates 
and $\Lambda$ (a wave vector, for instance, see section 4).

The most general solution of this equation can be written as a superposition of both complex and real
exponentials
\bee \label{15}
h_{\mu\nu}=\int\frac{d^4k}{(2\pi)^4}\delta(k^2-2\Lambda)\left(E_{\mu\nu}\cos{kx} + D_{\mu\nu}\sin{kx}
+\frac{\eta_{\mu\nu}}{4}\left( A\cosh{kx}+ B\sinh{kx}\right)\right)-\eta_{\mu\nu},
\ee
with $E_{\mu\nu}$ and $D_{\mu\nu}$ traceless, i.e. $E^\mu_{\,\mu}= D^\mu_{\,\mu}=0$. In the previous
expression $E_{\mu\nu}$, $D_{\mu\nu}$, $A$ and $B$ are in principle all independent 
functions of $k$ provided that the two following gauge conditions are met
\bee \label{16}
\int\frac{d^4k}{(2\pi)^4}\delta(k^2-2\Lambda) \left( k_\mu E_{\nu}^{\mu}\sin{kx}+\frac{k_{\nu}}{4}A\sinh{kx}
\right)=0
\ee
\bee \label{17}
\int\frac{d^4k}{(2\pi)^4}\delta(k^2-2\Lambda)\left(k_\mu D^{\mu}_{\nu}\cos{kx}-\frac{k_{\nu}}{4}B\cosh{kx}
\right)=0.
\ee
Clearly the integrands involved have to fall off sufficiently fast for large values of $k$ for 
the integrals to exist.

This solution has ten degrees of freedom to start with. Nine come from $E_{\mu\nu}$ and $D_{\mu\nu}$
after removal of the the trace. Another one comes from the coefficients $A, B$. Note that both $A$ and $B$ 
are needed to provide a full degree of freedom and likewise
for $E_{\mu\nu}$ and $D_{\mu\nu}$.
Using the gauge condition we can eliminate four of them, leaving six independent degrees of freedom. 
Unlike (\ref{13}), the above solution does not admit any residual gauge transformation to 
further eliminate degrees of freedom. 
Any attempt to perform a residual gauge transformation would take the solution `off-shell', 
i.e. the equations of motion would not be obeyed.

On the other hand we have to ensure that  
$h_{\mu\nu}<< 1 $; However, in general this does not eliminate any degree of freedom, it is just a 
requirement of the linearized theory. This translates in requiring the first term in the expansion of the
hyperbolic cosine to cancel the $-\eta_{\mu\nu}$ piece in (\ref{15}), or in other words
\bee \label{18}
\int\frac{d^4k}{(2\pi)^4}\delta(k^2-2\Lambda) A(k)= 4.
\ee

Since (\ref{15}) is the most general solution to the equations we must be able to recover 
the solutions in the previous section by performing an expansion in $\Lambda$. To do so we only have to 
choose the right form for $E_{\mu\nu}(k)$, $D_{\mu\nu}(k)$, $A(k)$ and $B(k)$.
As mentioned previously, to reach a Lorentz-covariant formulation such as (\ref{13}) in the
Lorenz gauge we can safely assume that $D_{\mu\nu}$ and $B$ are zero as the resulting
metric must satisfy $h_{\mu\nu}(x)= h_{\mu\nu}(-x)$, as discussed. In addition $A(k)$ can only be
a constant on Lorentz covariance grounds. We will take it to be $A(k)\equiv\frac{A'}{k^2}=\frac{A'}{2\Lambda}$. 
Also $E_{\mu\nu}$ needs to be a (traceless) Lorentz-covariant
tensor, namely 
$E_{\mu\nu}(k)\equiv\frac{E}{2\Lambda} \left(k_{\mu}k_{\nu}-\frac{\eta_{\mu\nu}}{2}\Lambda\right)$.  
The proportionality coefficient between $E$ and $A'$ comes from the gauge
condition (\ref{16}). Finally, as also indicated previously, the integrals require
a finite support to be well defined and this should be implemented in a Lorentz-invariant way too;
a sharp cut-off will be used below, although this is not crucial at all. Expanding (\ref{15}),
\bee \label{19}
\bal
h_{\mu\nu}=&\int\frac{d^4k}{(2\pi)^4}\delta(k^2-2\Lambda)\left(E_{\mu\nu}(k)\cos{kx}+\frac{\eta_{\mu\nu}}{4}
A(k)\cosh{kx}\right)-\eta_{\mu\nu}\\
=&\int\frac{d^4k}{(2\pi)^4}\delta(k^2-2\Lambda)\left(E_{\mu\nu}(k)\left(1-\frac{(k\cdot x)^2}{2}+\ldots\right)\right.\\
&\left. +\frac{\eta_{\mu\nu}}{4}
A(k)\left(1+\frac{(k\cdot x)^2}{2}+\ldots\right)\right)-\eta_{\mu\nu},\\
\eal
\ee
and using the definitions given above,
\bee
\bal \label{19.1}
h_{\mu\nu}\simeq &\int\frac{d^3\vec{k}}{(2\pi)^3}\frac{1}{2\sqrt{2\Lambda+
\vec{k}^2}}\left(\frac{E}{2\Lambda}\left(k_{\mu}k_{\nu}-
\frac{\eta_{\mu\nu}}{2}\Lambda\right)\left(1-\frac{(k\cdot x)^2}{2}\right)\right.\\
&\left. +\frac{\eta_{\mu\nu}}{4}\frac{A'}{2\Lambda}\left(1+\frac{(k\cdot x)^2}{2}\right)\right)-\eta_{\mu\nu}.\\
\eal
\ee
Now we introduce the cut-off, $\sqrt{2\Lambda}$. Already condition (\ref{18}) dictates 
the value for $A'=\frac{32\pi^2}{ C}$, where 
$C=\frac{1}{\Lambda}\int_{0}^{\sqrt{2\Lambda}}d|\vec{k}|\frac{\vec{k}^2}{\sqrt{2\Lambda+\vec{k}^2}}$. 
Then the solution reads
\bee \label{20}
\bal
h_{\mu\nu}\simeq &\int_ {0}^{\sqrt{2\Lambda}}\frac{d|\vec{k}|}{2\pi^2}\frac{\vec{k}^2}{2\sqrt{2\Lambda+
\vec{k}^2}}\left(-\frac{E}{2\Lambda}\left(k_{\mu}k_{\nu}-
\frac{\eta_{\mu\nu}}{2}\Lambda\right)\frac{(k\cdot x)^2}{2} 
+\frac{\eta_{\mu\nu}}{4}\frac{16\pi^2}{\Lambda C}\frac{(k\cdot x)^2}{2}\right)\\
=&\int_ {0}^{\sqrt{2\Lambda}}\frac{d|\vec{k}|}{2\pi^2}\frac{\vec{k}^2}{2\sqrt{2\Lambda+
\vec{k}^2}}\left(-E\left(\frac{\Lambda}{24}\left(\eta_{\mu\nu}x^2+2x_{\mu}x_{\nu}\right)-
\frac{\Lambda}{16}\eta_{\mu\nu}x^2\right) +\eta_{\mu\nu}x^2\frac{\pi^2}{C}\right)\\
=&\frac{\Lambda C}{4\pi^2}\left(-E\left(\frac{\Lambda}{24}\left(\eta_{\mu\nu}x^2+
2x_{\mu}x_{\nu}\right)-\frac{\Lambda}{16}\eta_{\mu\nu}x^2\right) +\eta_{\mu\nu}x^2\frac{\pi^2}{C}\right).\\
\eal
\ee
The value of $E$ is fixed via the gauge condition (\ref{16}) to $E=-\frac{16\pi^2}{3C\Lambda}$, leaving 
the perturbation in the form
\bee \label{21}
h_{\mu\nu}\simeq\frac{\Lambda}{9}\left(x_{\mu}x_{\nu}+2\eta_{\mu\nu}x^2\right),
\ee
which is precisely (\ref{13}).

\section{Wave-like solutions}
In this section we will finally investigate the effects 
of the cosmological constant in the propagation of GW in the appropriate coordinate system.

\subsection{Lowest order solutions}

We write $h_{\mu\nu}=h_{\mu\nu}^\Lambda + h_{\mu\nu}^W$. The term $h_{\mu\nu}^\Lambda$ 
is the solution we just found, $h_{\mu\nu}^W$ will be a perturbation on the metric induced 
by some source of GW. The same decomposition holds for the trace reversed metric $\tilde h_{\mu\nu}$. 
Waves are usually considered in the transverse traceless gauge\cite{TT}
\bee \label{14.2}
\tilde h^{W\mu}_{\,\mu} = h^{W\mu}_{\,\mu} =0, \qquad \partial_\mu h^{W\mu}_{\,\nu} 
= \partial_\mu\tilde h^{W\mu}_{\,\nu} =0.
\ee
This is compatible with the $\Lambda$-gauge condition as the r.h.s. of (\ref{lamg}) is
unchanged when considering $\tilde h_{\mu\nu}^\Lambda + \tilde h_{\mu\nu}^W$ provided that
(\ref{lamg}) is fulfilled by $h_{\mu\nu}^\Lambda$. This also makes clear that, 
at this order, the gauge condition involves the perturbation associated to the background 
and not the metric perturbation associated to a gravitational wave. 

Since the proper equations of motion in the Lorenz gauge at this order, neglecting $\mathcal{O}(\Lambda h_{\mu\nu})$, are just
$\Box h_{\mu\nu}= \Box h_{\mu\nu}^\Lambda + \Box h_{\mu\nu}^W = 0$, being the latter an
independent perturbation, it is obvious that 
\bee \label{fw}
\Box h_{\mu\nu}^W = 0,
\ee
and the gravitational wave solutions are in these coordinate systems functionally identical 
to those existing in flat space.

Note that because the $\Lambda h^{\Lambda}_{\mu\nu}$ has been neglected, the remaining residual
gauge invariance allows for a removal of four of the six degrees of freedom
in $h_{\mu\nu}^W$ and the analogy with wave propagation in Minkowski space is complete.

In the case of the lowest order equations the full solution of (\ref{10}) is
\bee \label{22}
h_{\mu\nu}=h_{\mu\nu}^{\Lambda}+h_{\mu\nu}^{W}=
\frac{\Lambda}{9}\left(x_{\mu}x_{\nu}+2\eta_{\mu\nu}x^2\right)+E_{\mu\nu}^{W}\cos{kx}+D_{\mu\nu}^{W}\sin{kx}
\ee
where $E^{W}=D^{W}=0$, $k_{\mu}E^{\mu W}_{\nu}=k_{\mu}D^{\mu W}_{\nu}=0$ and $k^2=0$. 

We want to see now how plane waves such as the ones in (\ref{22}) look like in the new coordinate system. 
Transformation (\ref{9.3}) acts both on the polarization tensors and on the arguments of the sine and cosine. 
For the polarization tensors we can always cut the expansion in $\Lambda$ and keep terms only up to a certain order. 
However, the transformation on the arguments yields terms of the type $Z^3w\Lambda$ which in general 
can be relevant. The sine and cosine can not be expanded, we have to transform the argument 
exactly; we shall later evaluate the error caused by retaining only the lowest order terms in the arguments. 

For the polarization tensors, since we transform them independently of the arguments, it is easy to 
see qualitatively what the corrections to the polarization tensors will be. On dimensional grounds 
alone, all corrections will be of order $\mathcal{O}(\sqrt{\Lambda}Z)$ or at most $\mathcal{O}(\Lambda Z^2)$, 
being these quantities in the region of validity of the approximation very small.

Nonetheless, the transformed wave-like solution to order $\sqrt{\Lambda}$ is
\bee \label{51}
\bal
h_{\mu\nu}^{W_{FRW}}=
&\begin{pmatrix}
  0 &  0 & 0 & 0  \\
  0 & E_{11}\left(1+2\sqrt{\frac{\Lambda}{3}}T\right) & E_{12}\left(1+2\sqrt{\frac{\Lambda}{3}}T\right) & 0 \\
 0  & E_{12}\left(1+2\sqrt{\frac{\Lambda}{3}}T\right) & -E_{11}\left(1+2\sqrt{\frac{\Lambda}{3}}T\right) &0 \\
    0   &0 &0 & 0 \\
    \end{pmatrix} \times \\
    &\cos{\left(w(T-Z)+w\sqrt{\frac{\Lambda}{3}}\left(\frac{Z^2}{2}-T Z\right)+\mathcal{O}(\Lambda)\right)} 
+\mathcal{O}(\Lambda)\\
&+\begin{pmatrix}
  0 &  0 &0 & 0  \\
 0& D_{11}\left(1+\sqrt{\frac{\Lambda}{3}}T\right) & D_{12}\left(1+\sqrt{\frac{\Lambda}{3}}T\right) & 0 \\
0  & D_{12}\left(1+\sqrt{\frac{\Lambda}{3}}T\right) & -D_{11}\left(1+\sqrt{\frac{\Lambda}{3}}T\right) &0 \\
    0   &0 &0 & 0 \\
    \end{pmatrix} \times\\
    &\sin{\left(w(T-Z)+w\sqrt{\frac{\Lambda}{3}}\left(\frac{Z^2}{2}-T Z\right)+\mathcal{O}(\Lambda)\right)}
    +\mathcal{O}(\Lambda)\\
\eal
\ee
The term $w(T-Z)$ dominates the argument of the trigonometric functions and it can be checked numerically 
that the error made by omitting terms of order $\Lambda$ or higher is $\leqslant 10^{-3}$ for the purposes 
of next section.

\subsection{Next-order solutions}

As we have argued before, it is not justified to neglect the term of order $\Lambda h_{\mu\nu}^W$ in this case, as
unlike for the case of the background, the magnitude of the two quantities is unrelated.
We can add a 
wave-like piece to the solution (\ref{15})
\bee \label{23}
\bal
h_{\mu\nu}=&h_{\mu\nu}^{\Lambda}+h_{\mu\nu}^{W}\\
=&\int\frac{d^4k}{(2\pi)^4}\delta(k^2-2\Lambda)\left(E_{\mu\nu}\cos{kx}+D_{\mu\nu}\sin{kx}+\frac{\eta_{\mu\nu}}{4}
\left(A\cosh{kx}+B\sinh{kx}\right)\right)-\eta_{\mu\nu}\\
&+E_{\mu\nu}^{W}\cos{kx}+D_{\mu\nu}^{W}\sin{kx}.
\eal
\ee
This will always be a solution of (\ref{lg}) and (\ref{6}) as long as 
$E^{W}=D^{W}=0$, $k_{\mu}E^{W\mu}_{\,\nu}=k_{\mu}D^{W\mu}_{\,\nu}=0$ and $k^2=2\Lambda$. 
However, now we are not allowed to perform any gauge transformation, at least at the next-order level. 
We can still use the gauge condition and the traceless condition to eliminate five degrees 
of freedom from the wave. We are left with a massive wave with five degrees of freedom. 
The polarization vectors of which, for a 
wave propagating in the $z$ direction ($k_{1}=k_{2}=0$), can be written as
\bee \label{24}
\bal
E_{\mu\nu}^{W}&=\begin{pmatrix}
    E_{00}  & \frac{\sqrt{w^2-2\Lambda}}{w}E_{13} & \frac{\sqrt{w^2-2\Lambda}}{w}E_{23} & \frac{w}{\sqrt{w^2-2\Lambda}}E_{00}  \\
    \frac{\sqrt{w^2-2\Lambda}}{w}E_{13}  & E_{11} & E_{12} & E_{13}  \\
  \frac{\sqrt{w^2-2\Lambda}}{w}E_{23}  & E_{12} & -E_{11}-E_{00}\frac{2\Lambda}{w^2-2\Lambda} & E_{23}  \\
   \frac{w}{\sqrt{w^2-2\Lambda}}E_{00}  & E_{13} &E_{23}  &   \frac{w^2}{w^2-2\Lambda} E_{00} \\
    \end{pmatrix}.
\eal
\ee
And a similar expression for $D^W_{\mu\nu}$.
At the exact level this is as far as one can go but in order to understand the meaning 
of these massive waves we turn again to an expansion in powers of $\Lambda$. We will proceed in two steps. 
First we expand the solution in powers of $\Lambda$ and collect terms order by order. Then, using the same 
reasoning in the equations of motion, we can use an approximate residual invariance to rewrite the 
polarization tensors as the usual GW in Minkowski space-time plus an order $\Lambda$ contribution 
with the extra degrees of freedom.

 The polarization vectors (\ref{24}) can then be written as
    \bee \label{25}
    \bal
  E_{\mu\nu}^{W}  &= \begin{pmatrix}
    E_{00}  & E_{13} & E_{23} &E_{00}  \\
   E_{13}  & E_{11} & E_{12} & E_{13}  \\
  E_{23}  & E_{12} & -E_{11} & E_{23}  \\
   E_{00}  & E_{13} &E_{23}  &   E_{00} \\
    \end{pmatrix}+\begin{pmatrix}
   0  &  -\frac{\Lambda}{w^2}E_{13} &  -\frac{\Lambda}{w^2}E_{23} &  \frac{\Lambda}{w^2}E_{00}  \\
     -\frac{\Lambda}{w^2}E_{13}  & 0 & 0 & 0  \\
   -\frac{\Lambda}{w^2}E_{23}  & 0 & -E_{00}\frac{2\Lambda}{w^2} & 0  \\
    \frac{\Lambda}{w^2}E_{00}  & 0 &0  &   \frac{2\Lambda}{w^2} E_{00} \\
    \end{pmatrix}+\mathcal{O}(\Lambda^2)\\
    &\equiv E_{\mu\nu}^{(0)}+E_{\mu\nu}^{(1)}+\mathcal{O}(\Lambda^2).\\
\eal
\ee
The same decomposition applies to $D_{\mu\nu}^{W}$. This expansion makes explicit the 
contributions of $\Lambda$ at a given order. We want to expand
\bee \label{26}
h_{\mu\nu}^{W}=h_{\mu\nu}^{(0)}+h_{\mu\nu}^{(1)}+\mathcal{O}(\Lambda^2),
\ee
where the superscript refers to the order in $\Lambda$.
The functions sine and cosine can also be expanded around a massless wave 
with coordinate-dependent amplitudes\cite{expansion} 
\begin{equation}\label{27}
\bal
h_{\mu\nu}^{W}&=E^{W}_{\mu\nu}\cos{kx}+D^{W}_{\mu\nu}\sin{kx}\\
&\simeq\left[\left(E^{W}_{\mu\nu}-\frac{\Lambda z}{w}D^{W}_{\mu\nu}\right)\cos{w(t-z)}
+\left(D^{W}_{\mu\nu}+\frac{\Lambda z}{w}E^{W}_{\mu\nu}\right)\sin{w(t-z)}\right]\\
\eal
\ee
or what is tantamount
\begin{equation}\label{28}
\bal
h_{\mu\nu}^{W}&=\left[\left(E_{\mu\nu}^{(0)}+E_{\mu\nu}^{(1)}-
\frac{\Lambda z}{w}D_{\mu\nu}^{(0)}\right)\cos{w(t-z)}\right.\\
&+\left.\left(D_{\mu\nu}^{(0)}+D_{\mu\nu}^{(1)}+\frac{\Lambda z}{w}E_{\mu\nu}^{(0)}\right)\sin{w(t-z)}\right]\\
&+\mathcal{O}(\Lambda^2)
\eal
\ee
We see that the massive wave we started with can be written at linear order in the cosmological constant in 
terms of a massless wave where all dependency in $\Lambda$ appears only through the polarization tensors
\bee \label{29}
h_{\mu\nu}^{W}=E_{\mu\nu}^{W}\cos{w(t-z)}+D_{\mu\nu}^{W}\sin{w(t-z)}+\mathcal{O}(\Lambda^2),
\ee
where $E_{\mu\nu}^{W}$  and $D_{\mu\nu}^{W}$ can be read from (\ref{28}).
The above is a valid solution of $\Box h_{\mu\nu}^{W}+2\Lambda h_{\mu\nu}^{W}=0$ only to 
order $\Lambda$ (included), which means we can expand the equations of motion to the same 
order without loss of validity
\bee \label{31}
\Box h_{\mu\nu}^{(0)}+\Box h_{\mu\nu}^{(1)}+2\Lambda h_{\mu\nu}^{(0)}+\mathcal{O}(\Lambda^2)=0
\ee
Now we can split the problem and solve order by order
\bee \label{32}
\bal
\Box h_{\mu\nu}^{(0)}=0\\
\Box h_{\mu\nu}^{(1)}+2\Lambda h_{\mu\nu}^{(0)}=0
\eal
\ee
Due to the fact that (\ref{31}) is not exact, the solution to it can admit a residual gauge transformation 
that will take the solution `off-shell' some order beyond the order we consider. For the transformed solution
\bee \label{33}
\bal
\Box h_{\mu\nu}^{\prime (0)}=0\\
\Box h_{\mu\nu}^{\prime (1)}+2\Lambda h_{\mu\nu}^{\prime (0)}=0.
\eal
\ee
The first equation in (\ref{33}) is analogous to (\ref{fw}), i.e. residual transformations 
on $h_{\mu\nu}^{(0)}$ are not restricted. To order zero we obtain GW analogous to the ones in flat space (in 
the present set of coordinates, that is). But in this case the transformation propagates to the 
following order through the second equation in (\ref{33}) 
making necessary to find the transformed $h_{\mu\nu}^{\prime (1)}$. 

It is not difficult to see that the following polarization tensor fulfills the necessary requirements
of tracelessness as well as the gauge condition ($k_{\mu}E^{W\mu}_{\,\nu}=k_{\mu}D^{W\mu}_{\,\nu}=0$)
\bee \label{40}
\bal
E_{\mu\nu}^{W}=&\begin{pmatrix}
   \frac{\Lambda}{w^2} E_{00} & -  \frac{\Lambda}{w^2} E_{13} & - \frac{\Lambda}{w^2}E_{23} & \frac{\Lambda}{w^2}E_{00}  \\
  -  \frac{\Lambda}{w^2} E_{13} & E_{11}-\frac{\Lambda z}{w}D_{11} & E_{12}-\frac{\Lambda z}{w}D_{12} & -  \frac{\Lambda}{w^2} E_{13} \\
  - \frac{\Lambda}{w^2}E_{23}  & E_{12}-\frac{\Lambda z}{w}D_{12} & -E_{11}+\frac{\Lambda z}{w}D_{11} &-  \frac{\Lambda}{w^2} E_{23} \\
     \frac{\Lambda}{w^2}E_{00}   &-  \frac{\Lambda}{w^2} E_{13} &-  \frac{\Lambda}{w^2} E_{23} &  \frac{\Lambda}{w^2}E_{00} \\
    \end{pmatrix}.
\eal
\ee
$D_{\mu\nu}$ is similarly obtained from (\ref{28}).
Notice the presence of the usual components (of ${\cal O}(1)$) in the 
polarization tensor in the $x,y$ entries of the metric.
 
To this order in $\Lambda$ we obtain massless 
waves with coordinate-dependent modified amplitudes which depend on $\Lambda$. 
We can see that the extra degrees of freedom due to the form of the 
linearized equations of motion for non-zero $\Lambda$  will only couple to matter fields proportionally to
$\Lambda$ thanks to the coupling $h^W_{\mu\nu} T^{\mu\nu}$
and thus will be irrelevant in practice. 

\subsection{Transformed next-order solutions}
Now we are ready to apply the series of coordinate transformations (\ref{42}, \ref{44}, \ref{46}, \ref{9.3}) 
to the wave-like solution (\ref{29}) that we found in the previous subsection in order to obtain 
a physical expression in FRW coordinates. Recall the waves in the general Lorenz gauge read 
\bee \label{50}
\bal
h_{\mu\nu}^{W}=E_{\mu\nu}^{W}(\Lambda ,z)\cos{w(t-z)}+D_{\mu\nu}^{W}(\Lambda ,z)\sin{w(t-z)},
\eal
\ee
where $E_{\mu\nu}^{W}$ can be read off from (\ref{40}). From (\ref{50}) it is clear the only modification 
with respect to the plane waves of the lower order is in the polarization tensors, being already of order $\Lambda$. 
This suggests that all the new modifications to order $\Lambda$ of the next-order waves are due to 
the change of coordinates. Explicitly the transformed waves to order $\Lambda$ read
\bee \label{51.1}
\bal
h_{\mu\nu}^{W_{FRW}}=
&\left[\begin{pmatrix}
   \frac{\Lambda}{w^2} E_{00} &  -  \frac{\Lambda}{w^2} E_{13} & - \frac{\Lambda}{w^2}E_{23} &  \frac{\Lambda}{w^2}E_{00}  \\
  -  \frac{\Lambda}{w^2} E_{13} & E_{11}-\frac{\Lambda Z}{w}D_{11} & E_{12}-\frac{\Lambda Z}{w}D_{12} & -  \frac{\Lambda}{w^2} E_{13} \\
 -  \frac{\Lambda}{w^2} E_{23}  & E_{12}-\frac{\Lambda Z}{w}D_{12} & -E_{11}+\frac{\Lambda Z}{w}D_{11} &-  \frac{\Lambda}{w^2} E_{23} \\
      \frac{\Lambda}{w^2} E_{00}   &-  \frac{\Lambda}{w^2} E_{13} &-  \frac{\Lambda}{w^2} E_{23} &   \frac{\Lambda}{w^2} E_{00} \\
    \end{pmatrix}\right. +\\
   & \left. \begin{pmatrix}
  0& 0 &0 & 0  \\
  0 & E_{11}\left(2\sqrt{\frac{\Lambda}{3}}T+\frac{2\Lambda}{9}T^2+\frac{5\Lambda}{18}Z^2\right) & E_{12}\left(2\sqrt{\frac{\Lambda}{3}}T+\frac{2\Lambda}{9}T^2+\frac{5\Lambda}{18}Z^2\right) & 0 \\
 0  & E_{12}\left(2\sqrt{\frac{\Lambda}{3}}T+\frac{2\Lambda}{9}T^2+\frac{5\Lambda}{18}Z^2\right) & -E_{11}\left(2\sqrt{\frac{\Lambda}{3}}T+\frac{2\Lambda}{9}T^2+\frac{5\Lambda}{18}Z^2\right) &0 \\
      0   &0 &0 &  0 \\
    \end{pmatrix}+\mathcal{O}(\Lambda^{3/2})\right]  \times \\
    &\cos{\left(w(T-Z)+w\sqrt{\frac{\Lambda}{3}}\left(\frac{Z^2}{2}-T Z\right)-\frac{1}{18} w \Lambda  \left(T^3+T^2 Z-5 T Z^2+2 Z^3\right)+\mathcal{O}(\Lambda^{3/2})\right)} 
\\
&+\left[\begin{pmatrix}
   \frac{\Lambda}{w^2} D_{00} &  -  \frac{\Lambda}{w^2} D_{13} & - \frac{\Lambda}{w^2}D_{23} &  \frac{\Lambda}{w^2}D_{00}  \\
  -  \frac{\Lambda}{w^2} D_{13} & D_{11}+\frac{\Lambda Z}{w}E_{11} & D_{12}+\frac{\Lambda Z}{w}E_{12} & -  \frac{\Lambda}{w^2} D_{13} \\
 -  \frac{\Lambda}{w^2} D_{23}  & D_{12}+\frac{\Lambda Z}{w}E_{12} & -D_{11}-\frac{\Lambda Z}{w}E_{11} &-  \frac{\Lambda}{w^2} D_{23} \\
      \frac{\Lambda}{w^2} D_{00}   &-  \frac{\Lambda}{w^2} D_{13} &-  \frac{\Lambda}{w^2} D_{23} &   \frac{\Lambda}{w^2} D_{00} \\
    \end{pmatrix}\right. +\\
   & \left. \begin{pmatrix}
  0& 0 &0 & 0  \\
  0 & D_{11}\left(2\sqrt{\frac{\Lambda}{3}}T+\frac{2\Lambda}{9}T^2+\frac{5\Lambda}{18}Z^2\right) & D_{12}\left(2\sqrt{\frac{\Lambda}{3}}T+\frac{2\Lambda}{9}T^2+\frac{5\Lambda}{18}Z^2\right) & 0 \\
 0  & D_{12}\left(2\sqrt{\frac{\Lambda}{3}}T+\frac{2\Lambda}{9}T^2+\frac{5\Lambda}{18}Z^2\right) & -D_{11}\left(2\sqrt{\frac{\Lambda}{3}}T+\frac{2\Lambda}{9}T^2+\frac{5\Lambda}{18}Z^2\right) &0 \\
      0   &0 &0 &  0 \\
    \end{pmatrix}+\mathcal{O}(\Lambda^{3/2})\right]  \times \\
    &\sin{\left(w(T-Z)+w\sqrt{\frac{\Lambda}{3}}\left(\frac{Z^2}{2}-T Z\right)-\frac{1}{18} w \Lambda  \left(T^3+T^2 Z-5 T Z^2+2 Z^3\right)+\mathcal{O}(\Lambda^{3/2})\right)}.
    \\
\eal
\ee

\section{Detectability}
Let us now do some order-of magnitude estimates to evaluate the effect of the corrections
induced by $\Lambda\neq 0$ on the propagation of gravitational waves. 

For the polarization tensors we have not attempted to derive the $\Lambda$-order corrections in full detail, 
although this is possible, because already the most relevant correction, i.e. $\sqrt{\Lambda}Z E_{\mu\nu}^{(0)}$, 
has to be some orders of magnitude smaller than $E_{\mu\nu}^{(0)}$ for the approximation to be valid. 
For example for a coordinate value of the order of a typical distance to a supernova, $10^{23}$ m, 
the quantity $\sqrt{\Lambda }Z\sim 10^{-3}$ ($\Lambda\sim 10^{-52}$ m$^{-2}$ $\sim 10^{-35}$ s$^{-2}$). 
This already means a small correction to an amplitude that has so far escaped detection and which presumably 
will not be measured with sufficient precision to discern the effect of the $\Lambda$-order effects in the foreseable future.
However, conceptually it is an interesting result.

It is more interesting to work out the corrections to the dispersion relation for (\ref{51}). As previously, let 
us consider waves that propagate in the $Z$ direction and are monochromatic. The maxima of the
wave will be reached when
\bee \label{52}
\bal
w(T-Z)+w\sqrt{\frac{\Lambda}{3}}\left(\frac{Z^2}{2}-T Z\right)=n\pi,
\eal
\ee
or
\bee \label{53}
\bal
Z_{\text{max}}(n,T)=T-\frac{n\pi}{w}-\frac{T^2}{2}\sqrt{\frac{\Lambda}{3}}+\frac{n^2\pi^2}{2w^2}\sqrt{\frac{\Lambda}{3}}.
\eal
\ee
From (\ref{53}) we can also calculate the phase velocity of the wave which is defined as
\bee \label{53.1}
v_{p}(T)\equiv\frac{dZ_{\text{max}}}{dT}=1-T\sqrt{\frac{\Lambda}{3}}+\mathcal{O}(\Lambda).
\ee
We see that in comoving coordinates the phase velocity is smaller than 1. This does not mean that the waves slow down. 
We can calculate the velocity in `ruler' distance. For a fixed time we have
\bee \label{53.2}
\bal
-dl^2=&-\left(1+T\sqrt{\frac{\Lambda}{3}}\right)dZ^2\\
\frac{dl}{dT}=&\frac{d}{dT}\left[\left(1+T\sqrt{\frac{\Lambda}{3}}\right)dZ_{\text{max}}\right]=1.
\eal
\ee

It is also interesting to rewrite the trigonometric functions of the wave defining 
$w_{\text{eff}}(Z)\equiv w\left(1-Z\sqrt{\frac{\Lambda}{3}}\right)$
\bee \label{54}
\cos\left[Tw\left(1-Z\sqrt{\frac{\Lambda}{3}}\right)-Zw\left(1-Z\sqrt{\frac{\Lambda}{3}}\right)\right]=\cos{w_{\text{eff}}(T-Z)}.
\ee
Note that the transformed wave corresponds to a usual wave with an effective frequency dependent on the coordinate $Z$. The wave becomes red-shifted as it propagates away from the source.

To see explicitly the effect of $\Lambda$ in the propagation of a wave described in comoving coordinates we plot (Figure 1) one 
of the $h_{++}$ components of the wave for a given instant ($T=0$ for simplicity). A wave with a physical frequency ranging $10^{3}\text{Hz}<w <10^{-10}\text{Hz}$ cannot be practically plotted in the relevant $Z$-range. To see the effect in a few cycles we take $w=4\cdot 10^{-16}$Hz, which does not affect the overall magnitude of the correction. We plot the wave for $\Lambda=10^{-52}m^{-2}$ and for $\Lambda=10^{-51}m^{-2}$ to assess the influence of $\Lambda$ on the wave propagation.
Then we plot $h_{++}\sim\left(1+\frac{5}{9}\Lambda Z^2\right)\cos\left[-Zw\left(1-Z\sqrt{\frac{\Lambda}{3}}\right)\right]$.
\begin{figure}[htbp]
\begin{center}
\includegraphics[scale=0.8]{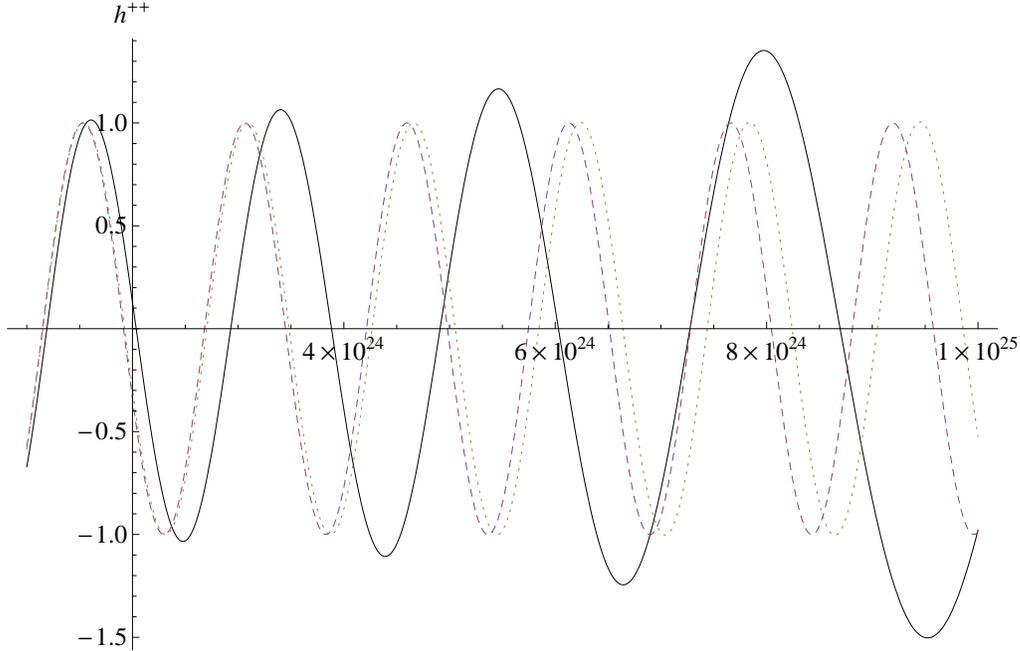}
\end{center}
\caption{Dependency of the amplitude and wave-length on the coordinate distance $Z$ (expressed in meters) for a constant value of $T$ and for different values of $\Lambda$: The dashed line corresponds to $\Lambda=0$, the dotted line to $\Lambda=10^{-52}m^{-2}$ and the solid line to $\Lambda=10^{-51}m^{-2}$.}
\end{figure}

From these results we can already draw some conclusions. The genuine corrections due to 
the mass-like term in (\ref{6}) remain unchanged in the transformed waves if we cut the expansion 
to order $\mathcal{O}(\Lambda)$. Moreover they are of order $\frac{\Lambda Z}{w}$, which is in practice 
irrelevant unless the value of $\Lambda$ is much greater than the current value. However, transformation (\ref{9.3}) 
induces modifications to the wave, both in the amplitude and the phase, of order $\sqrt{\Lambda}$ 
and $\Lambda$. This modifications result in a simulatneous increase  of the wave-length and of 
the amplitude with the coordinate $Z$. As shown in Figure 1, the most interesting region 
for detection would be that of events (supernovae and pulsars for example) happening at a distance $Z\sim 10^{23}-10^{25}$m away, 
for which the correction $\sqrt{\frac{\Lambda }{3}}Z\sim 10^{-1}-10^{-3}$ is not negligible and is 
well within the validity range of the approximation. In fact to have this type of correction into account
seems probably essential to properly account for the measurements of this type of phenomena in pulsar 
arrays.

\section{Summary}
The purpose of this work was to investigate the effect of the cosmological constant in the propagation 
of gravitational waves in a linearized theory of Gravity. The presence of $\Lambda$ leads unavoidably to the 
curvature of the background space-time in which the waves propagate. Within the linearized approximation
(which is the only framework where one can properly speak of `waves') this leads to a decomposition
$g_{\mu\nu}\simeq  \eta_{\mu\nu} + h_{\mu\nu}^\Lambda + h_{\mu\nu}^W$, including a modification
of the background (corresponding to the curvature) and a wave-like perturbation.

To see the way the propagation of the waves is affected, one 
has first to understand the implications that the different coordinate choices (gauge choices) have in the 
resolution of the equations of motion as well as the importance of the terms of different order retained in the linearization.
One is free to choose any particular gauge to solve the equations, however since the linearized Einstein 
equations are not invariant under general coordinate transformations their form will depend on the gauge choice.
We argue that the above procedure of linearization is consistent in some coordinate systems but not in others.
In particular, it is inconsistent to linearize the equations in the familiar Friedmann-Robertson-Walker cosmological coordinates (the metric only depends on time). 

Einstein equations can however be consistently linearized in Schwarzschild-de Sitter coordinates; then
$h_{\mu\nu}^{\Lambda}$ corresponds to a linearized version of the SdS metric, expanded to first order in $\Lambda$.
This metric can be easily modified to fulfill the Lorenz gauge condition. In this particular gauge, i.e. 
in this particular choice of coordinates, the analysis of gravitational waves follows a pattern very similar
to the one in Minkowski space-time. In the case where the $\Lambda h_{\mu\nu}$ term is dropped the residual gauge freedom of the Lorenz gauge allows for the removal of four additional degrees of freedom in the general solution, leaving 
the wave-like component with the usual two physical degrees of freedom of waves propagating in flat space-time. 

On the contrary, if the term $\Lambda h_{\mu\nu}$ is retained in the equations of motion the situation changes. Even in the Lorenz gauge the invariance under residual gauge transformations is lost. Again it is not hard to find the most general solution to the linearized equations composed of a background and a wave-like components. 
We prove the background solution to be consistent with the result previously found if $\Lambda$ is small. Since there 
is no residual invariance, the wave-like solution has to be interpreted as a massive wave with five degrees of freedom 
(the gauge condition and the trace condition amount to five constraints). However, we can make use of the approximate residual invariance at the leading order in $\Lambda$ to
rewrite the solution as massless gravitational waves with position-dependent modified amplitudes 
that change very slowly given the current values of $\Lambda$. There are only two ${\cal O}(1)$ polarizations; the remaining degrees of freedom (up to the five independent ones required for a massive spin two wave) are of ${\cal O}(\Lambda)$ and couple extremely weakly to matter sources.

Finally, one has to transform these solutions to the physically significant FRW coordinates in order to extract observable consequences. At this point modifications of ${\cal O}(\sqrt \Lambda)$ appear. Numerically these can be quite relevant for certain gravitational waves travelling from far away sources and the effect of $\Lambda$ can absolutely have a detectable impact on
pulsar timing arrays. \textsl{Waves are modified both in the phase and the amplitude; in cosmological coordinates
they are red-shifted in a prescribed way and the amplitude of plane waves grows as they move away from the source}.

\section*{Acknowledgements}
We acknowledge the financial support from the RTN ENRAGE and the research Grants FPA2007-66665, FPA2008-02878, 
FPA2010-20807, PROMETEO2008-004 and SGR2009SGR502. This research is supported by the Consolider CPAN
project. J.B. and D.E. would like to thank the CERN PH-TH Unit, where this research was initiated,
for the hospitality extended to them.
We thank J. Garcia-Bellido and R. Lapiedra for discussions on the subject.

\end{document}